\documentclass[prL,showpacs,twocolumn]{revtex4}

\usepackage{amssymb}

\newcommand{\be}{\begin{equation}}
\newcommand{\ee}{\end{equation}}
\newcommand{\ba}{\begin{eqnarray}}
\newcommand{\ea}{\end{eqnarray}}

\def\b{\beta}
\def\d{\delta}
\def\e{\epsilon}
\def\ve{\varepsilon}
\def\f{\phi}

\def\g{\gamma}
\def\h{\eta}
\def\j{\psi}

\def\m{\mu}
\def\n{\nu}

\def\r{\rho}

\def\x{\xi}

\def\D{\Delta}
\def\F{\Phi}
\def\G{\Gamma}

\def\O{\Omega}

\def\S{\Sigma}

\def\cf{{\cal F}}

\def\co{{\cal O}}

\def\cs{{\cal S}}

\newcommand{\ov}{\overline}
\newcommand{\uv}{\underline}

\newcommand{\wt}{\widetilde}
\newcommand{\wh}{\widehat}

\newcommand{\aand}{\;\;\;\mbox{and}\;\;\;}
\newcommand{\pa}{\partial}

\newcommand{\pari}{\stackrel{{P}}\longrightarrow}
\def\sl#1{\rlap{\hbox{$\mskip 1 mu /$}}#1}
\def\Sl#1{\rlap{\hbox{$\mskip 3 mu /$}}#1}
\def\SL#1{\rlap{\hbox{$\mskip 4 mu /$}}#1}
\def\I{\leavevmode\hbox{\small1\kern-3.8pt\normalsize1}}

\begin{document}
\title{The parity-preserving massive QED$_3$: vanishing $\beta$-function and no parity anomaly}
\author{O.M. Del Cima}
\email{oswaldo.delcima@ufv.br}
\affiliation{Universidade Federal de Vi\c cosa (UFV),\\
Departamento de F\'\i sica - Campus Universit\'ario,\\
Avenida Peter Henry Rolfs s/n - 36570-900 -
Vi\c cosa - MG - Brazil.}

\date{\today}
\begin{abstract}
\centerline{In honor of Prof. Raymond Stora (1930-2015)}
\vspace{0,25cm}

The parity-preserving massive QED$_3$ exhibits vanishing gauge coupling $\beta$-function and is parity and infrared anomaly free at all orders in perturbation theory. Parity is not an anomalous symmetry, even for the parity-preserving massive QED$_3$, in spite of some claims about the possibility of a perturbative parity breakdown, called parity anomaly. The proof is done by using the algebraic renormalization method, which is independent of any regularization scheme, based on general theorems of perturbative quantum field theory. 

\end{abstract}
\pacs{11.10.Gh 11.15.-q 11.15.Bt 11.15.Ex}
\maketitle


The quantum electrodynamics in three space-time dimensions (QED$_3$) has raised a great deal of 
interest since the precursor work by Deser, Jackiw and Templeton \cite{deser-jackiw-templeton} in 
view of a possible theoretical foundation for condensed matter phenomena, such as high-$T_{\rm c}$ 
superconductivity, quantum Hall effect and, more recently, graphene and topological insulators.
The massive and the massless QED$_3$ can exhibit interesting and subtle properties, namely superrenormalizability \cite{jackiw}, parity violation, topological gauge fields, anyons and the presence of infrared divergences. The massless QED$_3$ is ultraviolet and infrared perturbatively finite, infrared and parity anomaly free at all orders \cite{masslessQED3}, despite some statements found out in the literature that still support that parity could be broken even perturbatively, called parity anomaly, which has already been discarded \cite{masslessQED3,rao,delbourgo,pimentel}. The massless QED$_3$ is parity-even at the classical and quantum level (at least perturbatively), however, at the classical level, the massive QED$_3$ can be odd or even under parity symmetry. For the parity-even massive QED$_3$, if whether parity is 
a quantum symmetry or not, shall be definitely proved by using a renormalization method independent of any regularization scheme. The massive QED$_3$ has also been studied in details in many other physical configurations, namely, large gauge transformations, non-Abelian gauge groups, 
odd and even under parity, fermions families, compact space-times, space-times with boundaries, 
external fields and finite temperatures -- in all of these situations, the issues of parity 
breaking or parity preserving at the quantum level, renormalizability and finite temperature corrections, are quite non-trivial. Therefore, in those cases previously mentioned, the massive QED$_3$ shall exhibit distinct behaviours and properties \cite{non-trivial,CS} as compared to the case presented in this work.
 
The proof presented in this letter on the absence of parity and infrared anomaly, 
and the vanishing gauge coupling $\b$-function, in the parity-even massive QED$_3$, is based on
general theorems of perturbative quantum field theory \cite{qap,brs,pigsor,zimm}, where the 
Lowenstein-Zimmermann subtraction scheme in the framework of 
Bogoliubov-Parasiuk-Hepp-Zimmermann-Lowenstein (BPHZL) renormalization method \cite{zimm} is adopted. The former has to be introduced, owing to the presence of massless gauge field, so as to subtract infrared divergences that should arise from the ultraviolet subtractions.


The issue of the extension of parity-even massive QED$_3$ in the tree-approximation
to all orders in perturbation theory is organized according to two
independent parts. First, it is analysed the stability of the
classical action -- for the quantum theory, the stability corresponds
to the fact that the radiative corrections can be reabsorbed by a
redefinition of the initial parameters of the theory. Second,
it is computed all possible anomalies through an analysis of the Wess-Zumino
consistency condition, furthermore, it is checked if the possible breakings induced by
radiative corrections can be fine-tuned by a suitable choice of
local non-invariant counterterms. It shall be stressed that when massless fields are present, 
infrared divergences may appear from non-invariant counterterms, called infrared anomalies.

The gauge invariant action for the parity-preserving massive QED$_3$, with the gauge invariant
Lowenstein-Zimmermann (LZ) mass term added, is given by:
\ba
&&\S^{(s-1)}_{\rm inv}=\int{d^3 x} \biggl\{
-{1\over4}F^{\m\n}F_{\m\n} + i {\ov\j_+} {\Sl D} \j_+  + i {\ov\j_-} {\Sl D} \j_- + \nonumber\\
&&-~m({\ov\j_+}\j_+ - {\ov\j_-}\j_-) + 
\underbrace{{\frac\m2}(s-1)\e^{\m\r\n}A_\m\pa_\r A_\n}_{\rm LZ~mass~term}\biggr\}~,\label{inv}
\ea
where ${\SL D}\j_\pm\equiv(\sl\pa + ie \Sl{A})\j_\pm$, $e$ is a dimensionful
coupling constant with mass dimension $\frac12$, and $m$ is a mass parameter with mass dimension $1$. 
In action (\ref{inv}), $F_{\m\n}$ is the field strength for $A_\m$, 
$F_{\m\n}=\pa_\mu A_\nu - \pa_\n A_\m$, and, $\j_+$ and $\j_-$ are 
two kinds of fermions where the $\pm$ subscripts refer to their spin sign \cite{binegar}, also, 
the gamma matrices are $\gamma^\mu=(\sigma_z,i\sigma_x,i\sigma_y)$. The Lowenstein-Zimmermann parameter $s$ lies in the interval $0\le s\le1$ and plays the role of an additional subtraction variable (as the external momentum) in the BPHZL renormalization program, such that the parity-even massive QED$_3$ is
recovered for $s=1$.

In the BPHZL scheme a subtracted (finite) integrand, $R(p,k,s)$, is written in terms of the unsubtracted (divergent) one, $I(p,k,s)$, as 
\ba
R(p,k,s)&\!\!=\!\!&(1-t^{0}_{p,s-1})(1-t^{1}_{p,s})
I(p,k,s)\nonumber\\
&\!\!=\!\!&(1-t^{0}_{p,s-1}-t^{1}_{p,s}+t^{0}_{p,s-1}t^{1}_{p,s})
I(p,k,s)~, \nonumber
\ea  
where $t^{d}_{x,y}$ is the Taylor series about $x=y=0$ 
to order $d$ if $d\geq 0$. Thus, since the Lowenstein-Zimmermann mass term presented in (\ref{inv}) 
breaks parity, by assuming $s=1$, a subtracted integrand, $R(p,k,s)$, reads
\ba
R(p,k,1)=\underbrace{I(p,k,1)}_{\rm parity-even}-\underbrace{I(0,k,1)}_{\rm parity-even}-
\underbrace{p^\r\frac{\pa}{\pa p^\r}I(0,k,0)}_{\rm parity-odd~terms}~. \nonumber
\ea 

In order to quantize the model, represented by the action (\ref{inv}), a parity-even gauge-fixing action, $\S_{\rm gf}$, is added: 
\be
\S_{\rm gf}=\int{d^3 x}
\left\{b\pa^\m A_\m + {\x\over2}b^2 + {\ov c}\square c \right\}~,\label{gf}
\ee
together with a parity-even action term, $\S_{\rm ext}$, coupling the non-linear Becchi-Rouet-Stora (BRS) transformations to external sources:
\ba
\S_{\rm ext}&\!\!=\!\!&\int{d^3 x}
\bigl\{\ov\O_+ s\j_+ - \ov\O_- s\j_- + \nonumber\\
&\!\!-\!\!& s\ov\j_+ \O_+ + s\ov\j_- \O_-\bigr\}~.\label{ext}
\ea

The BRS transformations are given by:
\ba
&&s\j_+=ic\j_+~,~~s\ov\j_+=-ic\ov\j_+~,\nonumber\\
&&s\j_-=ic\j_-~,~~s\ov\j_-=-ic\ov\j_-~,\nonumber\\
&&sA_\mu=-{1\over e}\pa_\m c~,~~ sc=0~,\nonumber\\
&&s{\ov c}={1\over e}b~~,~~sb=0~, \label{BRS}
\ea
where $c$ is the ghost, ${\ov c}$ is the antighost and $b$ is the Lautrup-Nakanishi field \cite{lautrup-nakanishi}, playing the role of the Lagrange multiplier field. In spite of been massless, since the Faddeev-Popov ghosts are free fields, they decouple, therefore, no Lowenstein-Zimmermann mass term has to be introduced for them.

The complete action, $\S^{(s-1)}$, reads
\be
\S^{(s-1)}=\S^{(s-1)}_{\rm inv}+\S_{\rm gf}+\S_{\rm ext}~,\label{total}
\ee
in such a way that the parity-preserving massive QED$_3$ is recovered taking $s=1$, 
$\S\equiv\S^{(s-1)}|_{s=1}$.

By switching off the coupling constant ($e$) and taking the free part of the action, 
$\S^{(s-1)}_{\rm inv}+\S_{\rm gf}$ ((\ref{inv}) and (\ref{gf})), the tree-level propagators in momenta space, 
for all the fields, read: 
\ba
&&\D_{++}(k)=i\frac{\sl{k}+m}{k^2-m^2}~,~~\D_{--}(k)=i\frac{\sl{k}-m}{k^2-m^2}~,\label{propk++--}\\
&&\D^{\m\n}_{AA}(k,s)=
-i\biggl\{ \frac{1}{k^2-\m^2(s-1)^2}\biggl(\h^{\m\n}-\frac{k^\m k^\n}{k^2}\biggr) + \nonumber\\
&\!\!+\!\!& i\frac{\m(s-1)}{k^2[k^2-\m^2(s-1)^2]}\e^{\m\r\n}k_\r
+ \frac{\x}{k^2}\frac{k^\m k^\n}{k^2} \biggr\}~, \label{propkAA}\\
&&\D^\m_{Ab}(k)=\frac{k^\m}{k^2}~,~~\D_{bb}(k)=0~, \label{propkAbbb}\\
&&\D_{{\ov c}c}(k)=-i\frac{1}{k^2}~.\label{propkcbc}
\ea
At this moment, in order to establish the ultraviolet (UV) and infrared (IR) dimensions of any fields, 
$X$ and $Y$, we make use of the UV and IR asymptotical behaviour of their propagator, $\D_{XY}(k,s)$, $d_{XY}$ and $r_{XY}$, respectively: 
\ba
&&d_{XY}={\ov{\rm deg}}_{(k,s)}\D_{XY}(k,s)~, \\ 
&&r_{XY}={\uv{\rm deg}}_{(k,s-1)}\D_{XY}(k,s)~,
\ea
where the upper degree ${\ov{\rm deg}}_{(k,s)}$ gives the asymptotic power 
for $(k,s)\rightarrow \infty$ whereas the lower degree ${\uv{\rm deg}}_{(k,s-1)}$ 
gives the asymptotic power for $(k,s-1)\rightarrow 0$. The UV ($d$) and IR ($r$) 
dimensions of the fields, $X$ and $Y$, are chosen to fulfill the 
following inequalities:
\be
d_X + d_Y \geq 3 + d_{XY} \aand r_X + r_Y \leq 3 + r_{XY}~. \label{uv-ir}
\ee

In order to fix the UV and IR dimensions of the spinor fields $\j_+$ and $\j_-$, and the vector field $A_\m$, 
use has been made of the propagators, (\ref{propk++--}) and (\ref{propkAA}) together with the conditions 
(\ref{uv-ir}), then, the following relations stem:
\ba
 d_{\pm\pm}&\!\!=\!\!&-1~ \Rightarrow~ 2d_\pm\geq 2 \rightarrow d_\pm=1~, \label{d+-}\\
 r_{\pm\pm}&\!\!=\!\!&0~ \Rightarrow~ 2r_\pm\leq 3 \rightarrow r_\pm=\frac{3}{2}~; \label{r+-}\\
 d_{AA}&\!\!=\!\!&-2~ \Rightarrow~ 2d_A\geq 1 \rightarrow d_A=\frac{1}{2}~, \label{dA}\\
 r_{AA}&\!\!=\!\!&-2~ \Rightarrow~ 2r_A\leq 1 \rightarrow r_A=\frac{1}{2}~. \label{rA}
\ea
From the propagators (\ref{propkAbbb}) and the conditions, (\ref{uv-ir}), (\ref{dA}) and (\ref{rA}), it can fixed the UV and IR dimensions of the Lautrup-Nakanishi field $b$ as follows:
\ba
d_{Ab}&\!\!=\!\!&-1~ \Rightarrow~ d_A + d_b \geq 2 ~,~ d_A=\frac{1}{2} \rightarrow d_b=\frac{3}{2}~, \\
r_{Ab}&\!\!=\!\!&-1~ \Rightarrow~ r_A + r_b \leq 2 ~,~ r_A=\frac{1}{2} \rightarrow r_b=\frac{3}{2}~.
\ea
The dimensions (UV and IR) of the Faddeev-Popov ghost ($c$) and antighost (${\bar c}$) are fixed, by considering the propagators (\ref{propkcbc}), such that:
\ba
d_{{\bar c}c}&\!\!=\!\!&-2~ \Rightarrow~ d_c + d_{{\bar c}} \geq 1 ~, \\
r_{{\bar c}c}&\!\!=\!\!&-2~ \Rightarrow~ r_c + r_{{\bar c}} \leq 1 ~.
\ea 
Also, assuming that the BRS operator $s$ (\ref{BRS}) is dimensionless and bearing in mind 
that the coupling constant $e$ has dimension $(\rm{mass})^{\frac{1}{2}}$, the UV and IR dimensions for the ghost and antighost result:
\be
d_c=0 \aand d_{\bar c}=1 ~;~~ r_c=0 \aand r_{\bar c}=1 ~.
\ee 
Finally, from the action of the antifields (\ref{ext}), and the UV and IR dimensions of the fields fixed previously, it follows that:
\be
d_{\O_\pm}=2 \aand r_{\O_\pm}=\frac{3}{2} ~.
\ee

In summary, the UV ($d$) and IR ($r$) dimensions -- which are those involved in the Lowenstein-Zimmermann subtraction scheme \cite{zimm} -- as well as the ghost numbers ($\F\Pi$) and the Grassmann parity ($GP$) of all fields are collected in Table \ref{table1}. Notice that the statistics is defined as follows. The integer spin fields with odd ghost number, as well as, the half integer spin fields with even ghost number anticommute among themselves. However, the other fields commute with the formers and also among themselves.

\begin{table}
\begin{center}
\begin{tabular}{|c||c|c|c|c|c|c|c|c|c|c|}
\hline
    &$A_\mu$ &$\j_+$ &$\j_-$ &$c$ &${\ov c}$ &$b$ &$\O_+$ &$\O_-$ &$s-1$ &$s$ \\
\hline\hline
$d$ &${1/2}$ &1 &1 &0 &1 &${3/2}$ &2 &2 &1 &1 \\
\hline
$r$ &${1/2}$ &${3/2}$ &${3/2}$ &0 &1 &${3/2}$ &${3/2}$ &${3/2}$ &1 &0 \\
\hline
$\F\Pi$&0 &0 &0 &1 &$-1$ &0 &$-1$ &$-1$ &0 &0 \\
\hline
$GP$&0 &1 &1 &1 &1 &0 &0 &0 &0 &0 \\
\hline
\end{tabular}
\end{center}
\caption[]{UV ($d$) and IR ($r$) dimensions, ghost number ($\F\Pi$) and Grassmann parity ($GP$).}\label{table1}
\end{table}

The BRS invariance of the action is expressed in a functional way by the Slavnov-Taylor identity
\be
\cs(\S^{(s-1)})=0~,\label{slavnovident}
\ee
where the Slavnov-Taylor operator $\cs$ is defined, acting on an arbitrary functional $\cf$, by
\ba
\cs(\cf)&\!\!=\!\!&\int{d^3 x} \biggl\{-{1\over e}{\pa}^\mu c {\d\cf\over\d A^\mu} + {1\over e}b {\d\cf\over\d {\ov c}} + \nonumber\\
&\!\!+\!\!& {\d\cf\over\d \ov\O_+}{\d\cf\over\d \j_+} - 
{\d\cf\over\d \O_+}{\d\cf\over\d \ov\j_+} + \nonumber\\
&\!\!-\!\!& {\d\cf\over\d \ov\O_-}{\d\cf\over\d \j_-} + 
{\d\cf\over\d \O_-}{\d\cf\over\d \ov\j_-}\biggl\}~.\label{slavnov}
\ea
The corresponding linearized Slavnov-Taylor operator reads
\ba
\cs_\cf &\!\!=\!\!&\int{d^3 x} \biggl\{-{1\over e}{\pa}^\mu c {\d\over\d
A^\mu} + {1\over e}b {\d\over\d {\ov c}} + \nonumber\\
&\!\!+\!\!&{\d\cf\over\d \ov\O_+}{\d\over\d \j_+} + {\d\cf\over\d \j_+}{\d\over\d \ov\O_+} 
- {\d\cf\over\d \O_+}{\d\over\d \ov\j_+} - {\d\cf\over\d\ov\j_+}{\d\over\d \O_+} + \nonumber\\
&\!\!-\!\!&{\d\cf\over\d \ov\O_-}{\d\over\d \j_-} - {\d\cf\over\d \j_-}{\d\over\d \ov\O_-} + \nonumber\\
&\!\!+\!\!&{\d\cf\over\d \O_-}{\d\over\d \ov\j_-} + {\d\cf\over\d\ov\j_-}{\d\over\d \O_-}\biggl\}~.\label{slavnovlin}
\ea
The following nilpotency identities hold:
\ba
&&\cs_\cf\cs(\cf)=0~,~~\forall\cf~,\label{nilpot1} \\
&&\cs_\cf\cs_\cf=0~~{\mbox{if}}~~\cs(\cf)=0~. \label{nilpot3}
\ea
In particular, $(\cs_\S)^2=0$, since the action $\S^{(s-1)}$ obeys
the Slavnov-Taylor identity (\ref{slavnovident}). The operation of
$\cs_{\S}$ upon the fields and the external sources is given by
\ba
&&\cs_{\S}\f=s\f~,~~\f=\{\j_\pm,\ov\j_\pm,A_\m,c,{\ov c},b\}~,\nonumber\\
&&\cs_{\S}\O_+=-{\d\S^{(s-1)}\over\d\ov\j_+}~,~~\cs_{\S}\ov\O_+={\d\S^{(s-1)}\over\d\j_+}~,\nonumber\\
&&\cs_{\S}\O_-={\d\S^{(s-1)}\over\d\ov\j_-}~,~~\cs_{\S}\ov\O_-=-{\d\S^{(s-1)}\over\d\j_-}~. \label{operation1}
\ea
In addition to the Slavnov-Taylor identity (\ref{slavnovident}), the classical action $\S^{(s-1)}$ (\ref{total}) is characterized by the gauge condition, the ghost equation and the antighost equation:
\ba
{\d\S^{(s-1)}\over\d b}&\!\!=\!\!&\pa^\m A_\m + \x b~,\label{gaugecond}\\
{\d\S^{(s-1)}\over\d \ov c}&\!\!=\!\!&\square c~,\label{ghostcond}\\
-i{\d\S^{(s-1)}\over\d c}&\!\!=\!\!&i\square{\ov c} + \ov\O_+\j_+ +\ov\j_+\O_+ + \nonumber\\
&\!\!-\!\!& \ov\O_-\j_- - \ov\j_-\O_-~. \label{antighostcond}
\ea

The action $\S^{(s-1)}$ (\ref{total}) is invariant also with respect to the rigid symmetry
\be
W_{\rm rigid} \S^{(s-1)}=0~, \label{rigidcond}
\ee
where the Ward operator, $W_{\rm rigid}$, is defined by
\ba
&&W_{\rm rigid}=\nonumber\\
&\!\!=\!\!&\int{d^3 x}\biggl\{
\j_+{\d\over\d \j_+} - \ov\j_+{\d\over\d \ov\j_+} + \O_+{\d\over\d \O_+} - \ov\O_+{\d\over\d \ov\O_+} + \nonumber\\
&\!\!+\!\!& \j_-{\d\over\d \j_-} - \ov\j_-{\d\over\d \ov\j_-} + \O_-{\d\over\d \O_-} - 
\ov\O_-{\d\over\d \ov\O_-}
\biggr\}~. \label{wrigid}
\ea

The parity-preserving massive QED$_3$ ($s=1$) action, $\S^{(s-1)}|_{s=1}$, is invariant under parity ($P$), its action upon the fields and external sources is fixed as below:
\ba
x_\m & \pari & x_\m^P=(x_0,-x_1,x_2)~,\nonumber\\
\j_+ & \pari & \j_+^P=-i\g^1\j_-~,~\ov\j_+ \pari \ov\j_+^P=i\ov\j_-\g^1 ,\nonumber \\
\j_- & \pari & \j_-^P=-i\g^1\j_+~,~\ov\j_- \pari \ov\j_-^P=i\ov\j_+\g^1 ,\nonumber \\
A_\mu & \pari & A_\mu^P=(A_0,-A_1,A_2)~,\nonumber\\
\f & \pari & \f^P=\f~,~~\f=\{c,\bar c,b\}~,\nonumber\\
\O_+ & \pari & \O_+^P=-i\g^1\O_-~,~\ov\O_+ \pari \ov\O_+^P=i\ov\O_-\g^1 ,\nonumber \\
\O_- & \pari & \O_-^P=-i\g^1\O_+~,~\ov\O_- \pari \ov\O_-^P=i\ov\O_+\g^1 .
\label{xp}
\ea

In order to verify if the action in the tree-approximation ($\S^{(s-1)}$) is stable under radiative corrections, we perturb it by an arbitrary integrated local functional (counterterm) $\S^{c (s-1)}$, 
such that
\be
\wt\S^{(s-1)}=\S^{(s-1)}+\ve \S^{c (s-1)}~, \label{adef}
\ee
where $\ve$ is an infinitesimal parameter. The functional $\S^c\equiv\S^{c (s-1)}|_{s=1}$ has the same 
quantum numbers as the action in the tree-approximation at $s=1$.

The deformed action $\wt\S^{(s-1)}$ must still obey all the conditions presented above, henceforth, 
$\S^{c (s-1)}$ is subjected to the following set of constraints:
\ba
&&\cs_{\S}\S^{c (s-1)}=0~, \label{stabcond}\\
&&{\d\S^{c (s-1)}\over{\d b}}={\d\S^{c (s-1)}\over{\d{\ov c}}}={\d\S^{c (s-1)}\over{\d c}}=0~, \label{cond}\\
&&W_{\rm rigid} \S^{c (s-1)}=0~. \label{crigidcond}
\ea

The most general invariant counterterm $\S^{c (s-1)}$ -- the most general field polynomial -- with UV and IR dimensions bounded by $d\le3$ and $r\ge3$, with ghost number zero and fulfilling the conditions displayed in Eqs.(\ref{stabcond})-(\ref{crigidcond}), reads:
\ba
\S^{c (s-1)}&\!\!=\!\!&\int{d^3 x} \bigl\{
\alpha_1 i{\ov\j}_+{\Sl D}\j_+ +  \alpha_2 i{\ov\j}_-{\Sl D}\j_- +\nonumber\\
&\!\!+\!\!& \alpha_3 {\ov\j}_+\j_+ +  \alpha_4 {\ov\j}_-\j_- +\nonumber\\
&\!\!+\!\!& \alpha_5 F^{\m\n}F_{\m\n} + \alpha_6 \e^{\m\r\n}A_\m \pa_\r A_\n \bigr\}~. \label{finalcount}
\ea
where $\alpha_i$ ($i=1,\ldots,6$) are, in principle, arbitrary parameters. However, there are other restrictions owing to the superrenormalizability of the theory and its parity invariance -- the parity-even massive QED$_3$ recovered for $s=1$. On account of the superrenormalizability, the coupling constant-dependent power-counting formula~\cite{CS,YMCS-BFYM} is given by:
\be
\bordermatrix{& \cr                   
&\d(\g) \cr
&\r(\g)  }
 = 3 - \sum\limits_\F 
\bordermatrix{& \cr                   
&d_\F \cr
&r_\F  } N_\F - \frac 12 N_e~, \label{power}
\ee
for the UV ($\d(\g)$) and IR ($\r(\g)$) degrees of divergence of a 1-particle irreducible Feynman graph,
$\g$. Here $N_\F$ is the number of external lines of $\g$ corresponding to the field $\F$, $d_\F$ and $r_\F$ are the UV and IR dimensions of $\F$, respectively, as given in Table \ref{table1}, and $N_e$ is the power of the coupling constant $e$ in the integral corresponding to the diagram $\g$. Due to the fact that the counterterms are generated by loop graphs, they are at least of order two in the coupling constant ($e$). Consequently, the effective UV and IR dimensions of the counterterm $\S^{c (s-1)}$ are bounded by $d\le2$ and $r\ge2$, then, $\alpha_1=\alpha_2=\alpha_5=0$. Furthermore, the counterterm 
$\S^c\equiv\S^{c (s-1)}|_{s=1}$ is parity invariant, yielding that $\alpha_6=0$ and $\alpha_3=-\alpha_4=\alpha$. Finally, it can be concluded that the counterterm results as
\ba
\S^c\equiv\S^{c (s-1)}|_{s=1}&\!\!=\!\!&\int{d^3 x}\{\alpha({\ov\j}_+\j_+ - {\ov\j}_-\j_-)\}~, \nonumber\\
&\!\!=\!\!&z_m m\frac{\pa}{\pa m}\S~,
\label{ct}
\ea
where $z_m$ is an arbitrary parameter (as $\alpha$ is, $z_m=-\frac{\alpha}{m}$), and $\S\equiv\S^{(s-1)}|_{s=1}$. The counterterm (\ref{ct}) shows that, {\it a priori}, only the mass parameter $m$ can get radiative 
corrections. This means that the $\beta_e$-function related to the gauge coupling constant ($e$) is vanishing 
($\beta_e=0$) to all orders of perturbation theory, so as the anomalous dimensions of the fields.


Owing to the fact that classical stability does not imply the possibility of extending the theory to the quantum level, it still lacks to show the absence of gauge anomaly, infrared anomaly and the claimed parity anomaly. This result, combined with the previous one (\ref{ct}), completes the proof of vanishing gauge coupling $\beta$-function and the absence of infrared and parity anomaly in parity-even massive QED$_3$ at all orders in perturbation theory.

At the quantum level the vertex functional, $\G^{(s-1)}$, which coincides with the classical action, 
$\S^{(s-1)}$ (\ref{total}), at $0$th order in $\hbar$,
\be
\G^{(s-1)}=\S^{(s-1)} + {\co}(\hbar)~,\label{vertex}
\ee
has to satisfy the same constraints as the classical action does, namely  
Eqs.(\ref{gaugecond})-(\ref{rigidcond}).

In accordance with the Quantum Action Principle {\cite{qap,pigsor}}, the
Slavnov-Taylor identity (\ref{slavnovident}) gets a quantum breaking:
\be
\cs(\G^{(s-1)})|_{s=1}=\D \cdot \G^{(s-1)}|_{s=1} = \D + {\co}(\hbar \D)~,
\label{slavnovbreak}
\ee
where $\D\equiv\D|_{s=1}$ is an integrated local functional, taken at $s=1$,
with ghost number $1$ and UV and IR dimensions bounded by $d\le\frac72$ and
$r\ge3$.

The nilpotency identity ({\ref{nilpot1}) together with
\be
\cs_{\G}=\cs_{\S} + {\co}(\hbar)~,
\ee
implies the following consistency condition for the breaking $\D$:
\be
\cs_{\S}\D=0~,\label{breakcond1}
\ee
and beyond that, $\D$ also satisfies the constraints:
\be
{\d\D\over\d b}={\d\D\over\d\ov c}=\int d^3x \frac{\d\D}{\d c}=
W_{\rm rigid}\D=0~.\label{breakcond5}
\ee

The Wess-Zumino consistency condition (\ref{breakcond1}) constitutes a
cohomology problem in the sector of ghost number one.
Its solution can always be written as a sum of a trivial cocycle
$\cs_{\S}{\wh\D}^{(0)}$, where ${\wh\D}^{(0)}$ has ghost number $0$,
and of nontrivial elements belonging to the cohomology of $\cs_{\S}$
(\ref{slavnovlin}) in the sector of ghost number one:
\be
\D^{(1)} = {\wh\D}^{(1)} + \cs_{\S}{\wh\D}^{(0)}~.\label{breaksplit}
\ee

It shall be stressed that there still remains a possible parity violation
at the quantum level induced by parity-odd noninvariant counterterms. Due
to the fact that the Lowenstein-Zimmermann subtraction method breaks
parity during the intermediary steps, the Slavnov-Taylor identity breaking,
$\D^{(1)}$, is not necessarily parity invariant. In any case, $\D^{(1)}$
must satisfy the conditions imposed by (\ref{breakcond1}) and (\ref{breakcond5}).
The trivial cocycle $\cs_{\S}{\wh\D}^{(0)}$ can be absorbed into the vertex
functional $\G^{(s-1)}$ as a noninvariant integrated local counterterm,
$-{\wh\D}^{(0)}$. On the other hand, a nonzero ${\wh\D}^{(1)}$ would 
represent an anomaly. If by chance, there exist any parity-odd ${\wh\D}_{\rm odd}^{(0)}$,
a parity anomaly would be present induced by the noninvariant counterterm, $-{\wh\D}_{\rm odd}^{(0)}$.

Taking into account the Slavnov-Taylor operator $\cs_{\S}$ (\ref{slavnovlin}) and the quantum breaking (\ref{slavnovbreak}), it results that the breaking $\D^{(1)}$ exhibits UV and IR dimensions bounded by
$d\le{7\over2}$ and $r\geq3$. Nevertheless, being an effect of the radiative corrections, the insertion $\D^{(1)}$ possesses a factor $e^2$ at least, then its effective UV and IR dimensions are bounded by $d\le{5\over2}$ and $r\geq2$, respectively.

From the antighost equation: 
\be
\int d^3x \frac{\d\D^{(1)}}{\d c}=0~,
\ee
it follows that $\D^{(1)}$ can be written as 
\be
\D^{(1)} = \int{d^3 x}~{\cal K}_\mu\pa^\m c~,
\ee
where ${\cal K}_\mu$ is a rank-$1$ tensor with ghost number $0$, with UV and IR dimensions bounded by $d\le{3\over2}$ and $r\geq1$ (the ghost $c$ is dimensionless), respectively. The breaking $\D^{(1)}$ can be split into two pieces, which are even and odd under parity, by writing ${\cal K}_\mu$ as 
\be
{\cal K}_\mu = r_{\rm v} {\cal V}_\mu + r_{\rm p} {\cal P}_\mu~,
\ee 
in such a manner that ${\cal V}_\mu$ is a vector and ${\cal P}_\mu$ a pseudo-vector. 

Bearing in mind that ${\cal K}_\mu$ has its UV and IR dimensions bounded by $d\le{3\over2}$ and $r\geq1$, it can be concluded that there are no ${\cal V}_\mu$ satisfying these dimensional constraints 
and the conditions (\ref{breakcond1}) and (\ref{breakcond5}), therefore, $\{{\cal V}_\mu\}=\emptyset$, which means the absence of a parity-even Slavnov-Taylor breaking. However, still remains the odd sector represented by ${\cal P}_\mu$, and it follows that by a dimensional analysis a candidate for ${\cal P}_\mu$, which satisfies also the conditions (\ref{breakcond1}) and (\ref{breakcond5}), shows up:
\be
{\cal P}_\mu = {\wt F}_\m = \frac{1}{2}~\e_{\m\r\n}F^{\r\n}~.
\ee 
It turns out that there is only one parity-odd candidate, $\D_{\rm odd}^{(1)}$, which could be a parity anomaly, surviving all the constraints above:
\be
\D^{(1)} = \D_{\rm odd}^{(1)} = \frac{r_{\rm p}}{2} \int{d^3 x}~\e_{\m\r\n}F^{\r\n}\pa^\m c~,\label{oddpbreak}
\ee
however, integrating it by parts, leads that 
\be
\D^{(1)} = \D_{\rm odd}^{(1)} \equiv 0~.
\ee  
Hence it follows that, there is no radiative corrections to the insertion describing the breaking of the Slavnov-Taylor identity, $\{\D^{(1)}\}=\emptyset$, which means that there is no possible breaking to the Slavnov-Taylor identity, and neither parity is violated nor infrared anomaly stems by noninvariant counterterms that could be induced due to the Lowenstein-Zimmermann subtraction method, which breaks parity at the intermediary stages of the IR subtractions.


In conclusion, the parity-preserving massive QED$_3$ exhibits vanishing gauge coupling 
$\b$-function ($\b_e=0$), vanishing anomalous dimensions of all the fields ($\g_\F=0$), and besides that, is infrared and parity anomaly free at all orders in perturbation theory. In fact, the latter is a by-product of superrenormalizability and absence of parity-odd noninvariant couterterms that could be induced by the IR divergences subtractions which break parity -- there is no Chern-Simons term radiatively induced at any order as some claims found out in the literature. It shall be stressed that the algebraic renormalization method  does not involve any regularization scheme, nor any particular diagrammatic calculation, and is based on 
general theorems of perturbative quantum field theory.


O.M.D.C. dedicates this work to his father (Oswaldo Del Cima, {\it in memoriam}), mother (Victoria M. Del Cima, {\it in memoriam}), daughter (Vittoria) and son (Enzo). He also thanks the referee 
for useful comments and suggestions.

\end{document}